\documentclass[sigconf]{acmart}

\AtBeginDocument{%
	\providecommand\BibTeX{{%
			\normalfont B\kern-0.5em{\scshape i\kern-0.25em b}\kern-0.8em\TeX}}}

\acmConference[]{ArXiv}{2021}{}
\acmBooktitle{}
\acmPrice{}
\acmISBN{}
\usepackage{amsmath}
\usepackage{amsfonts}
\usepackage{graphicx}
\usepackage{float}
\usepackage{algpseudocode}
\usepackage{algorithm}
\usepackage{algorithmicx}
\usepackage{url}
\usepackage{caption}
\usepackage{color}
\usepackage{paralist}

\newtheorem{example}{Example}

\newcommand{\market}{\mathcal{M}}

\newcommand{\buyer}{b}
\newcommand{\buyerset}{\mathcal{B}}
\newcommand{\seller}{s}
\newcommand{\sellerset}{\mathcal{S}}
\newcommand{\player}{g}
\newcommand{\playerset}{\mathcal{G}}
\newcommand{\productset}{\mathcal{D}}
\newcommand{\product}{d}
\newcommand{\costset}{C}
\newcommand{\cost}{c}
\newcommand{\winset}{W}
\newcommand{\win}{w}
\newcommand{\valuationset}{V}
\newcommand{\valuation}{v}
\newcommand{\budget}{L}
\newcommand{\horizon}{T}

\newcommand{\simname}{\textsf{OSOUM}}

\title{\simname{} Framework for Trading Data Research}
\author{Gregory Goren, Roee Shraga, Alexander Tuisov}
\authornote{authors contributed equally to this research.}
\email{{gregory.goren,shraga89, alexandt}@campus.technion.ac.il}
\affiliation{
	\institution{Technion -- Israel Institute of Technology}
	\city{Haifa}
	\country{Israel}
}

\begin{document}
	\begin{abstract}
    In the last decades, data have become a cornerstone component in many business decisions, and copious resources are being poured into production and acquisition of the high-quality data. This emerging market possesses unique features, and thus came under the spotlight for the stakeholders and researchers alike. In this work, we aspire to provide the community with a set of tools for making business decisions, as well as analysis of markets behaving according to certain rules. We supply, to the best of our knowledge, the first open source simulation platform, termed \textbf{O}pen \textbf{SOU}rce \textbf{M}arket Simulator (\simname) to analyze trading markets and specifically data markets. We also describe and implement a specific data market model, consisting of two types of agents: sellers who own various datasets available for acquisition, and buyers searching for relevant and beneficial datasets for purchase. The current simulation treats data as an infinite supply product. Yet, other market settings may be easily implemented using \simname. Although commercial frameworks, intended for handling data markets, already exist, we provide a free and extensive end-to-end research tool for simulating possible behavior for both buyers and sellers participating in (data) markets.
	\end{abstract}
	\maketitle
	
	\section{Introduction}\label{sec:introduction}
	In the last decades, data have become a cornerstone component in many business decisions, and copious resources are being poured into production and acquisition of the high-quality data. This emerging market possesses unique features, and thus came under the spotlight for the stakeholders and researchers alike. Consider the following case:


\begin{example}\label{ex:intro}
A new content provider is stepping into the entertainment industry, providing streaming media and video on demand. To understand its target customers, the provider chooses to use the power of data. Typical relevant data sources may include competing content providers and complementary industries that may hold valuable information regarding potential customers. In terms of a data market, the new provider has to decide which of the relevant datasets should it purchase, when, and at what cost. Buying all the relevant datasets is usually impossible as the initial budget usually is not sufficient, and even if so, postponing the purchase may result in a drop in buying prices. However, having a consistent benefit from the retained datasets may encourage the buyer to buy the datasets early, even if the supply is infinite, making timing an important ingredient. 
\end{example}
In this work, we aspire to provide the community with a set of tools for making business decisions, as well as analysis of markets behaving according to certain rules. We supply, to the best of our knowledge, the first open source simulation platform, termed \textbf{O}pen \textbf{SOU}rce \textbf{M}arket Simulator (\simname) to analyze trading markets and specifically data markets. We also describe and implement a specific data market model, consisting of two types of agents: sellers who own various datasets available for acquisition, and buyers searching for relevant and beneficial datasets for purchase. The current simulation treats data as an infinite supply product. Yet, other market settings may be easily implemented using \simname.  \par
Although commercial frameworks, intended for handling data markets, already exist (\emph{e.g.,}~\cite{Dawex,onaudience,Qlik,tase,snowflake}), we provide a free and extensive end-to-end research tool for simulating possible behavior for both buyers and sellers participating in (data) markets. Our main contribution can be summarized as follows:

\begin{compactitem}
	\item We provide a publicly available novel market simulation, including a convenient method of adding custom behaviors for both the sellers and buyers, as well as a wide variety of possible contracts between the parties.
	\item We implement several baseline approaches to the following tasks: 
	\begin{enumerate}
	    \item Estimating market prices for data products, 
	    \item Choosing subset of these products to be procured, based on price prediction and personal valuations
	    \item Maximizing profit by purchase timing w.r.t. market prices
	\end{enumerate}
	
\end{compactitem}

We begin by describing \simname{} (Section~\ref{sec:sim}) and discussing related work on data markets (Section~\ref{sec:related}). Then, we provide an \simname{} proof-of-concept handling data trading (Section~\ref{sec:datamarket}).

	\section{The OSOUM Framework}\label{sec:sim}
	We now describe our market simulation framework, \simname{}, which is publicly available at \url{https://github.com/shraga89/RGA}. Using \simname{}, researchers can flexibly implement different market scenarios in various settings, run simulations, analyze markets behaviour and more. \simname{} is composed of \emph{players}, \emph{products}, and \emph{simulator}. 

We distinguish between two main types of players: buyers and sellers. Generally, both share most functionalities, including price setting, strategy, budget, \emph{etc.} In addition, each player stores a history of transactions to make flexible use of past interactions for future decisions. \simname{} allows players to employ different strategies with regard to any decision made, may it be market price prediction, choosing which goods to sell/purchase, price setting or something else. Strategies can be either picked from an existing pool or implemented using \simname{}.

The simulation supports both consumable and non-consumable products, as well as items with finite and infinite supplies, e.g. physical and digital goods respectively.  

\simname{} also allows different market rules and various types of contracts between players. For example, a simple supply and demand market simulation (finite supply) without auctions is implemented as a simple interaction between seller and buyers (randomly assigned pairs at each timestamp). In this setting a transaction occurs if the buying price exceeds the selling price; accordingly, the budgets of the buyers and sellers are updated as well as the inventory. \simname{} also supports selling via various types of auctions and provides the tools for easy addition of other forms of contracts and financial instruments. 
	\section{Related Work}\label{sec:related}
Before we dive into our proof-of-concept dealing with data markets, we now describe related work in the area.
A variety of approaches and frameworks have addressed data markets. From a commercial perspective, an abundant of frameworks exists, \emph{e.g., } Dawex~\cite{Dawex}, Onaudience~\cite{onaudience}, Qlik~\cite{Qlik}, Tase~\cite{tase} and snowflake's data exchange~\cite{snowflake}). These frameworks act as mediators and enable transactions between competing sellers and buyers. Using our suggested data market simulation, buyers and sellers willing to use these frameworks will have the opportunity to test and analyze their strategies before committing to a real-world environment.  

Originated in economics, where trading data has been ongoing for more than 30 years now~\cite{varian1997versioning}, research into data markets has been a focus for other disciplines as well, such as data management~\cite{koutris2015query,chawla2019revenue,chen2019towards}. In game theory, the focus is on the theory of creating algorithmic solutions specifically tailored for data-like products, and taking their unique traits (infinite replicability, combinatorial value \emph{etc.}) into account \cite{agarwal2019marketplace, acemoglu2019too}. In this work we focus on a market setup simulation for be used for other researchers. In addition, while other works use the characteristics of data (size, features, use in ML framework, cleanness, \emph{etc.}), we provide a proof-of-concept for an infinite supply market, not limited to data products. 



	\section{Simulating a Data Market using OSOUM}\label{sec:datamarket}
	

In this section we give a concrete example of a market model that can be faithfully simulated and investigated using \simname.


\subsection{A Data Market Model}\label{sec:market_model}

A \emph{data market} $\market$ is a composition of three sets of entities, namely \emph{buyers} $\buyerset$, \emph{sellers} $\sellerset$, and a \emph{mediator}. We shall refer to the first two entities also as the \emph{players} $\playerset = \buyerset\cup\sellerset$ in the market. Players may be individuals or groups (\emph{e.g.,} companies or organizations) interested in trading datasets. We use $\productset$ to denote the set of authorized datasets in $\market$. Usually, a dataset $\product \in \productset$ is associated with a domain and several of other properties that represent it including \emph{e.g.}, the features it contains. Each buyer $\buyer\in\buyerset$ is interested in a set of datasets $\productset_{\buyer}\subseteq\productset$ whereas each seller offers a set of products $\productset_{\seller}\subseteq\productset$. In addition, each player has a budget $\budget_{\player}$, which is updated according to the transactions of a player. The mediator is in charge of the transactions between the different players in the market.

Each player $\player\in\playerset$ has a different utility from a dataset and a different perceived value, according to which it can set prices. A market is a temporal ecosystem. We assume that the market has a finite horizon $\horizon$ and each interaction between the different entities takes place in a discrete timestamp $t<T$. Each timestamp $t$ includes a set of transactions supervised by the mediator.

Being a part of a data market, players must decide what is the value of a dataset, \emph{i.e.,} how profitable is a dataset for them. Establishing a value for a dataset (and information goods in general) is not easy~\cite{varian1997versioning,castro2020data}. The core challenge in a competing market is that the value of data is different for sellers and buyers. Sellers may price a dataset reflecting \emph{e.g.,} the effort in gathering the data, while buyers may choose the price they are willing to pay according to their profit expectation. We separate the price of a product from it valuation which are denoted as $p_{\player}(\product)$ and $v_{\player}(\product)$, respectively. 

A \emph{transaction} is when a buyer $\buyer$ acquires a dataset $\product$ from a seller $\seller$. The buyer pays $p(\product)$, $\budget_{\buyer}$ and $\budget_{\seller}$ are updated, and the buyer is no longer interested in buying $\product$. The seller is still willing to sell $\product$, as the inventory of a dataset is assumed to be unlimited. At the end of each timestamp, buyers and sellers update their datasets pricing.

\begin{example}
	Recall Example~\ref{ex:intro}. Our content provider is denoted by $\buyer1$. Let its relevant datasets be $\productset_{\buyer1} = \{\product_1, \product_2\}$, and its initial budget by $\budget_{\buyer1} = 10$. $\buyer1$ price estimations are $p_{\buyer1}(\product_1) = 8$ and $p_{\buyer1}(\product_2) = 4$ and its valuations are $p_{\buyer1}(\product_1) = 3$ and $p_{\buyer1}(\product_2) = 2$.\footnote{note that the valuations are acquired each timestamp, \emph{e.g.,} having $\product_1$ for 5 timestamps yields a value of 15} In this case, $\buyer1$ cannot aim to buy both $\product_1$ and $\product_2$ and has to choose just one of them. This decision may depend on the horizon and can be expressed as $2T-4>3T-8$. Then, for example, if $T<4$ then buying $\product_2$ is more beneficial, yielding \emph{e.g.,} a profit of 3 compared to 2 for $T=3$. However, for $T>4$ its more beneficial to buy $\product_1$, yielding \emph{e.g.,} a profit of 7 compared to 6 for $T=5$. If $T=4$, both will yield 4. 
\end{example}

In practice, it is not always possible to purchase all the profitable products because of either funds available being insufficient, or because of pricing mismatch between a buyer and a seller. So a choice must be made at each time point what are the datasets to be bought immediately. We call this choice \emph{datasets allocation} problem.



\subsection{Datasets Allocation Optimization Strategy}\label{sec:allocation}
The \emph{datasets allocation} problem is defined with respect to a single player $\player$ before timestamp $t$ in the market horizon $T$. Recall that a player $\player$ is associated with $\productset = \langle \product_1, \product_2, \dots, \product_n\rangle$ (set of relevant datasets) and $\budget$ (budget). In addition, a dataset $D_{i}$ has an overall valuation value $\valuation_i = (T-t)\cdot\valuation_{\player}(D_{i})$. 
At timestamp $0$, the set of datasets $\productset$, its corresponding valuations $\valuationset$ and an initial $\budget$ are set. While the valuation values of datasets stays constant throughout the horizon, the $\productset$ and $\budget$ change with respect to the interaction in the market. The available datasets at time $t$, $\productset^{t}$, changes when a player purchases a dataset and thus, $\productset^{t} \subseteq \productset^{t^{\prime}}, t^{\prime}<t$ and $\productset^{0} = \productset$. The budget $\budget^{t}$ changes with respect to the revenue of purchased products and costs.


At time $t$ the player has to select a subset of datasets $\bar{\productset} \subseteq \productset^{t-1}$ that maximizes her future revenues. To simplify the notation, we denote that number of datasets available at time $t$ as $m\leq n$. Let $\costset = \langle \cost_1, \cost_2, \dots, \cost_{m}\rangle, c_i\in {\rm I\!R}$ and $\winset = \langle \win_1, \win_2, \dots, \win_m\rangle, \win_{m}\in \{0,1\}$ represent a realization of costs and win indicators of the relevant datasets after timestamp $t$ has completed, respectively. In practice, a player has to allocate a subset of products at the beginning of time $t$. Thus, the player does not know the actual cost of datasets when the dataset allocation takes place. Accordingly, the player has to estimate the costs $\hat{\costset} = \langle \hat{\cost}_1, \hat{\cost}_2, \dots, \hat{\cost}_m\rangle$ and winning indicators $\hat{\winset} = \langle \hat{\win}_1, \hat{\win}_2, \dots, \hat{\win}_m\rangle$. Using these estimations, the \emph{datasets allocation} problem can be formalized as follows:

\begin{equation}\label{eq:main1}
\begin{aligned}
& \underset{X}{\text{maximize}} & & \sum_{i=1}^{m} (\valuation_{i} - \hat{\cost}_{i})\cdot\hat{\win}_{i}\cdot X_{i} \\
& \text{subject to} & & \sum_{i=1}^{m} \hat{\cost}_{i}\cdot\hat{\win}_{i}\cdot X_{i} \leq \budget \\
& & & X_{i} \in \{0,1\} \; i = 1, \ldots, m. & 
\end{aligned}
\end{equation}

\noindent Since $\hat{\win}_{i}\in \{0,1\}$, the size of $\productset^{t-1}$ can only decrease to the set of product she estimates she would win, \emph{i.e.,} $win(\productset^{t-1}) = \{p_{i}\in \productset^{t-1} | \hat{w}_{i} = 1\}$. Denoting the size of $win(\productset^{t-1})$ as $m_{win}$, we can modify the objective to be $\sum_{i=1}^{m_{win}} (\valuation_{i} - \hat{\cost}_{i})\cdot X_{i}$ and the constraints to be $\sum_{i=1}^{m_{win}} \hat{\cost}_{i}\cdot X_{i} \leq \budget$ and $X_{i} \in \{0,1\} \; i = 1, \ldots, m_{win}.$, respectively.

Exploiting the resemblance to the Knapsack problem~\cite{karp1975computational}, one can show that our problem is NP-hard (via a reduction) and use an out-of-the-box solver, \emph{e.g.,} Gurobi,\footnote{\url{https://www.gurobi.com/}} to (optimally) solve the problem.

\subsection{Simulating the Data Market} 
We present an example of data market simulation based on the model defined in Section~\ref{sec:market_model}. The data market was implemented as a part of \simname{}, and can effortlessly be extended to incorporate far more sophisticated techniques.

\subsubsection{Goods}
In this context, the only type of product is a dataset. As such, it stores the domain of the dataset, its size (in terms of examples and features) as well as other information. 

\subsubsection{Buyers}\label{sec:buyers}
Recall that buyers have different valuations for each dataset (Section~\ref{sec:market_model}) and employ different cost estimation strategies. One can restate the price prediction problem faced by a buyer as a time series analysis problem. For sake for demonstration, we next present some basic baseline buying techniques. We note that there exist far more sophisticated methods, \emph{e.g.,}~\cite{agarwal2019marketplace}, which we leave for future works. Let $p_1, p_2, \dots p_t$ be a sequence of market prices for a dataset, we predict the value of $p_{t+1}$ using the following methods : 
\begin{compactitem}
\item \textbf{last} price ($p_{t+1} = p_t$)
\item \textbf{mean} price ($p_{t+1} = \frac{1}{t}\sum_{i=1}^{t}p_i$)
\item \textbf{max}imum price ($p_{t+1} = \max\limits_{i=1,\dots,t} p_i$)
\item \textbf{min}imum price ($p_{t+1} = \min\limits_{i=1,\dots,t}p_i$)
\item linear \textbf{regression} $\left(p_{t+1} = f\left(p_1, p_2, \dots p_t\right)\right)$
\end{compactitem}




Using the predicted prices, a buyer has to choose which products to buy, since the buyers' budget is limited, and buying everything is rarely possible. Thus, a buyer picks the most valuable products as described in Problem~\ref{eq:main1}. In an auction-free market, the buyers do not compete among themselves for a dataset. Accordingly, acquiring a dataset depends solely on the interaction between the buyer and potential sellers. Thus, Problem~\ref{eq:main1} does not need to consider winning indicators, allowing us to use knapsack to solve the problem. 

\subsubsection{Sellers}

In our setting, we focus on competing buyers as they aim to maximize their utility. A buyer's strategy, however, may be strongly affected by sellers behavior; thus we cannot completely ignore this issue. To demonstrate the verity of sellers in such setting, we created a set of reasonable ad-hoc rules that dictate sellers' behavior, without any guarantee of optimality. The implemented sellers are as follows:
\begin{compactitem}
    \item \textbf{Adaptive seller}: increases price if she sold the product successfully, and lowers the price otherwise. 
    \item \textbf{Linear seller}: price is a linear function of time.
    \item Noisy variants of the sellers above, using Gaussian noise. 
\end{compactitem}

\subsubsection{Methodology}\label{sec:auction_free}

We created a proof-of-concept experiment based on an auction-free market. Let $\buyerset$ and $\sellerset$ be the buyers and sellers, respectively. Each player in the market behaves independently. For each dataset $\product$ available in the market, each $\buyer$ (such that $\product \in \productset_{\buyer}^{t}$) is randomly assigned to a seller $\seller$ (such that $\product \in \productset_{\seller}$). Then, we split into two cases:
\begin{itemize}
    \item \textbf{If $p_{\buyer}(\product) \geq p_{\seller}(\product)$:} the buyer obtains the dataset ($\productset_{\buyer}^{t} = \productset_{\buyer}^{t}\setminus \product$) and pays its selling price ($\budget_{\buyer} = \budget_{\buyer} - p_{\seller}(\product)$). 
    \item \textbf{If $p_{\buyer}(\product) < p_{\seller}(\product)$:} the buyer does not obtain the dataset and get allocated to a different seller.
\end{itemize}\noindent Note that in both cases the dataset remains available for sale. The process continues until no buyer is interested in acquiring this $\product$. During a timestamp the set prices remain constant. Once it terminates, each player has the opportunity to change its prices. For example, if a seller did not sell any instance of a dataset it may lower the price. 

\subsubsection{Example Results}
\begin{figure}
    \centering
    \includegraphics[width=8cm,height=8cm,keepaspectratio]{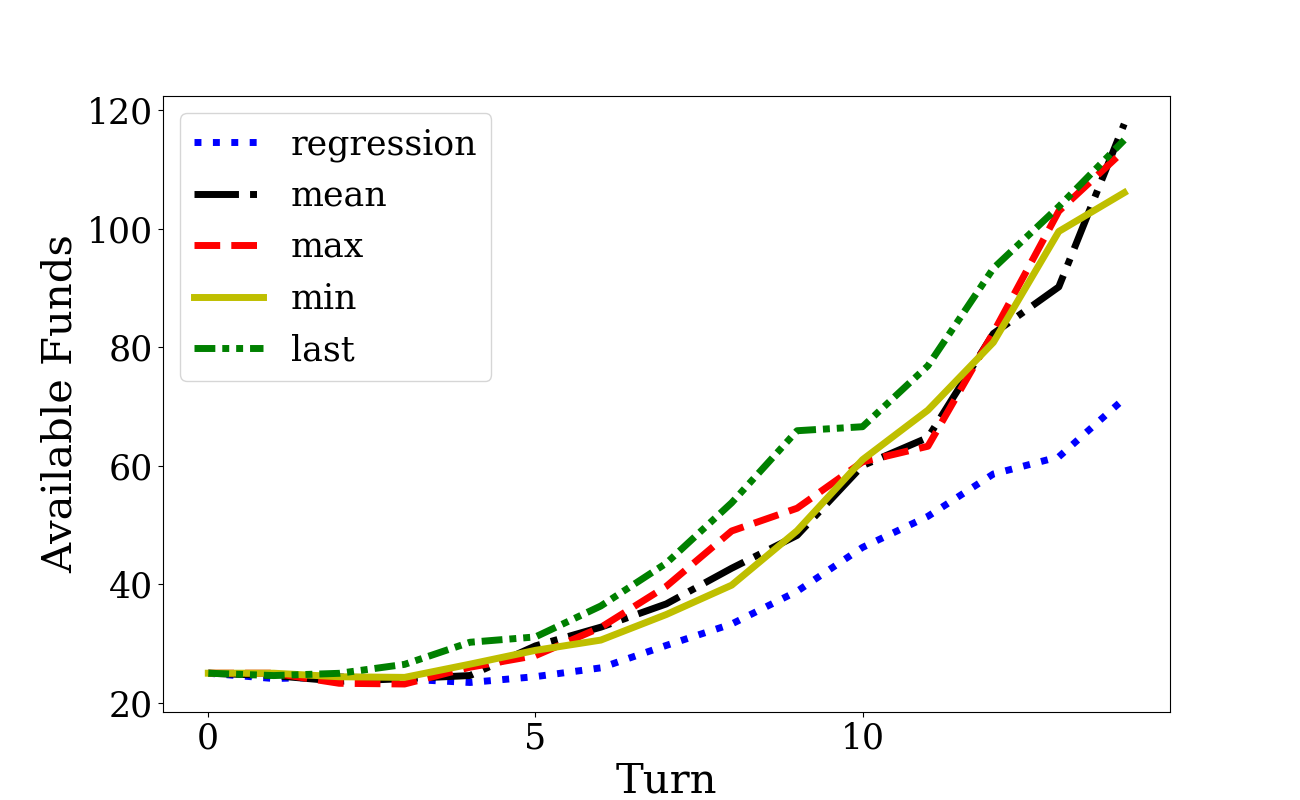}
    \caption{Available funds of buyers with different cost estimation strategies per turn. Values are averaged over 10 different simulations.}
    \label{fig:funds}
    \vspace{-3mm}
\end{figure}

Figure~\ref{fig:funds} presents a possible plot based on a simulation outcome, referring to five types of buyers (see Section~\ref{sec:buyers}). The y-axis represents the funds (or budget) available to each buyer at timestamp $t$. We do not provide a prior price estimation at $t=0$ and thus, the initial values of cost estimations are drawn randomly. In what follows, we average the values over 10 different simulation runs to minimize the randomization effect. We observe inferiority in the estimation performance of regression based buyer. This may be explained due by the non-linear behaviour of some of the sellers as well as the fact that a considerable amount of data is needed for a reasonable regression model. Last price buyer is able to obtain a judicious cost estimations right from the start and the costs do not change vastly between timestamps in our configuration.









    
    \balance
	\bibliographystyle{ACM-Reference-Format}
	\bibliography{refs}


\begin{thebibliography}{13}


\ifx \showCODEN    \undefined \def \showCODEN     #1{\unskip}     \fi
\ifx \showDOI      \undefined \def \showDOI       #1{#1}\fi
\ifx \showISBNx    \undefined \def \showISBNx     #1{\unskip}     \fi
\ifx \showISBNxiii \undefined \def \showISBNxiii  #1{\unskip}     \fi
\ifx \showISSN     \undefined \def \showISSN      #1{\unskip}     \fi
\ifx \showLCCN     \undefined \def \showLCCN      #1{\unskip}     \fi
\ifx \shownote     \undefined \def \shownote      #1{#1}          \fi
\ifx \showarticletitle \undefined \def \showarticletitle #1{#1}   \fi
\ifx \showURL      \undefined \def \showURL       {\relax}        \fi
\providecommand\bibfield[2]{#2}
\providecommand\bibinfo[2]{#2}
\providecommand\natexlab[1]{#1}
\providecommand\showeprint[2][]{arXiv:#2}

\bibitem[\protect\citeauthoryear{??}{Daw}{[n.d.]}]%
        {Dawex}
 \bibinfo{year}{[n.d.]}\natexlab{}.
\newblock \bibinfo{title}{Dawex}.
\newblock \bibinfo{howpublished}{\url{https://www.dawex.com/en/}}.
\newblock


\bibitem[\protect\citeauthoryear{??}{ona}{[n.d.]}]%
        {onaudience}
 \bibinfo{year}{[n.d.]}\natexlab{}.
\newblock \bibinfo{title}{OnAudience}.
\newblock \bibinfo{howpublished}{\url{https://www.onaudience.com/}}.
\newblock


\bibitem[\protect\citeauthoryear{??}{Qli}{[n.d.]}]%
        {Qlik}
 \bibinfo{year}{[n.d.]}\natexlab{}.
\newblock \bibinfo{title}{Qlik Datamarket}.
\newblock
  \bibinfo{howpublished}{\url{https://www.qlik.com/us/products/qlik-data-market}}.
\newblock


\bibitem[\protect\citeauthoryear{??}{sno}{[n.d.]}]%
        {snowflake}
 \bibinfo{year}{[n.d.]}\natexlab{}.
\newblock \bibinfo{title}{Snowflake Data Exchange}.
\newblock
  \bibinfo{howpublished}{\url{https://www.snowflake.com/data-exchange/}}.
\newblock


\bibitem[\protect\citeauthoryear{??}{tas}{[n.d.]}]%
        {tase}
 \bibinfo{year}{[n.d.]}\natexlab{}.
\newblock \bibinfo{title}{Tase}.
\newblock \bibinfo{howpublished}{\url{https://www.tase.co.il/en/market_data}}.
\newblock


\bibitem[\protect\citeauthoryear{Acemoglu, Makhdoumi, Malekian, and
  Ozdaglar}{Acemoglu et~al\mbox{.}}{2019}]%
        {acemoglu2019too}
\bibfield{author}{\bibinfo{person}{Daron Acemoglu}, \bibinfo{person}{Ali
  Makhdoumi}, \bibinfo{person}{Azarakhsh Malekian}, {and}
  \bibinfo{person}{Asuman Ozdaglar}.} \bibinfo{year}{2019}\natexlab{}.
\newblock \bibinfo{booktitle}{\emph{Too much data: Prices and inefficiencies in
  data markets}}.
\newblock \bibinfo{type}{{T}echnical {R}eport}. \bibinfo{institution}{National
  Bureau of Economic Research}.
\newblock


\bibitem[\protect\citeauthoryear{Agarwal, Dahleh, and Sarkar}{Agarwal
  et~al\mbox{.}}{2019}]%
        {agarwal2019marketplace}
\bibfield{author}{\bibinfo{person}{Anish Agarwal}, \bibinfo{person}{Munther
  Dahleh}, {and} \bibinfo{person}{Tuhin Sarkar}.}
  \bibinfo{year}{2019}\natexlab{}.
\newblock \showarticletitle{A marketplace for data: An algorithmic solution}.
  In \bibinfo{booktitle}{\emph{Proceedings of the 2019 ACM Conference on
  Economics and Computation}}. \bibinfo{pages}{701--726}.
\newblock


\bibitem[\protect\citeauthoryear{Castro~Fernandez, Subramaniam, and
  Franklin}{Castro~Fernandez et~al\mbox{.}}{2020}]%
        {castro2020data}
\bibfield{author}{\bibinfo{person}{Raul Castro~Fernandez},
  \bibinfo{person}{Pranav Subramaniam}, {and} \bibinfo{person}{Michael~J
  Franklin}.} \bibinfo{year}{2020}\natexlab{}.
\newblock \showarticletitle{Data Market Platforms: Trading Data Assets to Solve
  Data Problems [Vision Paper]}.
\newblock  (\bibinfo{year}{2020}).
\newblock


\bibitem[\protect\citeauthoryear{Chawla, Deep, Koutrisw, and Teng}{Chawla
  et~al\mbox{.}}{2019}]%
        {chawla2019revenue}
\bibfield{author}{\bibinfo{person}{Shuchi Chawla}, \bibinfo{person}{Shaleen
  Deep}, \bibinfo{person}{Paraschos Koutrisw}, {and} \bibinfo{person}{Yifeng
  Teng}.} \bibinfo{year}{2019}\natexlab{}.
\newblock \showarticletitle{Revenue maximization for query pricing}.
\newblock \bibinfo{journal}{\emph{Proceedings of the VLDB Endowment}}
  \bibinfo{volume}{13}, \bibinfo{number}{1} (\bibinfo{year}{2019}),
  \bibinfo{pages}{1--14}.
\newblock


\bibitem[\protect\citeauthoryear{Chen, Koutris, and Kumar}{Chen
  et~al\mbox{.}}{2019}]%
        {chen2019towards}
\bibfield{author}{\bibinfo{person}{Lingjiao Chen}, \bibinfo{person}{Paraschos
  Koutris}, {and} \bibinfo{person}{Arun Kumar}.}
  \bibinfo{year}{2019}\natexlab{}.
\newblock \showarticletitle{Towards Model-based Pricing for Machine Learning in
  a Data Marketplace}. In \bibinfo{booktitle}{\emph{Proceedings of the 2019
  International Conference on Management of Data}}.
  \bibinfo{pages}{1535--1552}.
\newblock


\bibitem[\protect\citeauthoryear{Karp}{Karp}{1975}]%
        {karp1975computational}
\bibfield{author}{\bibinfo{person}{Richard~M Karp}.}
  \bibinfo{year}{1975}\natexlab{}.
\newblock \showarticletitle{On the computational complexity of combinatorial
  problems}.
\newblock \bibinfo{journal}{\emph{Networks}} \bibinfo{volume}{5},
  \bibinfo{number}{1} (\bibinfo{year}{1975}), \bibinfo{pages}{45--68}.
\newblock


\bibitem[\protect\citeauthoryear{Koutris, Upadhyaya, Balazinska, Howe, and
  Suciu}{Koutris et~al\mbox{.}}{2015}]%
        {koutris2015query}
\bibfield{author}{\bibinfo{person}{Paraschos Koutris}, \bibinfo{person}{Prasang
  Upadhyaya}, \bibinfo{person}{Magdalena Balazinska}, \bibinfo{person}{Bill
  Howe}, {and} \bibinfo{person}{Dan Suciu}.} \bibinfo{year}{2015}\natexlab{}.
\newblock \showarticletitle{Query-based data pricing}.
\newblock \bibinfo{journal}{\emph{Journal of the ACM (JACM)}}
  \bibinfo{volume}{62}, \bibinfo{number}{5} (\bibinfo{year}{2015}),
  \bibinfo{pages}{1--44}.
\newblock


\bibitem[\protect\citeauthoryear{Varian}{Varian}{1997}]%
        {varian1997versioning}
\bibfield{author}{\bibinfo{person}{Hal~R Varian}.}
  \bibinfo{year}{1997}\natexlab{}.
\newblock \bibinfo{booktitle}{\emph{Versioning information goods}}.
\newblock \bibinfo{type}{{T}echnical {R}eport}.
\newblock


\end{thebibliography}
\end{document}